\documentclass[usenatbib, referee]{mn2e}
\usepackage{graphicx}
\usepackage{epsfig}
\usepackage{color}

\title[MSPs with wide orbits]
{Formation of millisecond pulsars with wide orbits}
\author[B. Wang et al.]
{Bo Wang,$^{\rm 1,2,3}$\thanks{E-mail: wangbo@ynao.ac.cn}
Dongdong Liu,$^{\rm 1,2,3}$\thanks{E-mail: liudongdong@ynao.ac.cn}  
Yunlang Guo,$^{\rm 4,5}$\thanks{E-mail: yunlang@nju.edu.cn}
Hailiang Chen,$^{\rm 1,2,3}$ 
\newauthor
Wenshi Tang,$^{\rm 4,5}$
Luhan Li$^{\rm 1,2,3}$
and Zhanwen Han$^{\rm 1,2,3}$ \\ 
$^1$Yunnan Observatories, Chinese Academy of Sciences, Kunming 650216, China\\
$^2$International Centre of Supernovae, Yunnan Key Laboratory, Kunming 650216, China\\
$^3$University of Chinese Academy of Sciences, Beijing 100049, China\\
$^4$School of Astronomy and Space Science, Nanjing University, Nanjing 210023, China\\
$^5$Key Laboratory of Modern Astronomy and Astrophysics, Nanjing University, Ministry of Education, Nanjing 210023, China}
\begin{document}
\date{Accepted. Received}
\pagerange{\pageref{firstpage}--\pageref{lastpage}} \pubyear{2024}
\maketitle

\label{firstpage}

\begin{abstract}
Millisecond pulsars (MSPs) are a kind of radio pulsars with short spin periods,  
playing a key role in many aspects of stellar astrophysics.   In recent years, 
some more MSPs with wide orbits ($>30\,\rm d$) have been discovered, but their origin is still highly unclear.
In the present work, according to an adiabatic power-law assumption for the mass-transfer  process, 
we carried out a large number of complete binary evolution computations for the formation of MSPs with wide orbits
through the iron core-collapse supernova (CCSN) channel, in which 
a  neutron star (NS)  originating from a CCSN accretes matter from a red-giant (RG) star and spun up to millisecond periods.
We found that  this channel can form the observed MSPs with wide orbits in the range of $30-1200\,{\rm d}$, in which 
the WD companions have masses in the range of $0.28-0.55\,\rm M_{\odot}$. 
We also found that almost all the observed MSPs  can be reproduced  
by this channel in the WD companion mass versus orbital period diagram. 
We estimate that the Galactic numbers of the resulting MSPs from the CCSN channel are in the  range of $\sim 4.8-8.5\times10^{5}$.
Compared with the accretion-induced collapse  channel,
the CCSN channel provides a main way to produce MSPs with wide orbits.

\end{abstract}


\section{Introduction} 

Millisecond pulsars (MSPs) are an important subclass of rotation powered radio pulsars 
with short spin pulse periods (usually $<30$\,ms) and weak surface magnetic fields 
($\sim10^{8}-10^{9}\,{\rm G}$),  about 80\% of which
are discovered in binaries  (see e.g. Manchester 2004; Lorimer 2008; Han et al. 2021).
As the end point of binary evolution, MSPs play a key role in many aspects of stellar astrophysics.  
The observed stellar and orbital properties of binary MSPs  can be used to constrain 
some important processes during binary evolution, 
such as the mass-accretion onto neutron stars (NSs), the common-envelope evolution and 
the angular momentum loss mechanisms,  etc (see Tauris 2011; D'Antona \& Tailo 2020;  Tauris \& van den Heuvel 2023). 
Meanwhile, the exotic evolutionary history and environments  combined with the extreme timing stability
makes MSPs excellent  astrophysics labs to study matter in extreme states (see Bhattacharyya \& Roy 2021). 

MSPs are believed to be formed from the evolution of  low-mass X-ray binaries (LMXBs), where NSs
were produced by iron core-collapse  supernovae (CCSNe) 
and have subsequently been spun up to millisecond periods through sufficient mass accretion from their companions,
called the standard recycling model (see Alpar et al. 1982; Chanmugam \& Brecher 1987; 
Bhattacharya  \& van den Heuvel 1991; Podsiadlowski, Rappaport \& Pfahl 2002).
The recycling model is now widely accepted for the formation of MSPs, but there remain 
many processes that are not well understood, such as the mass-accretion physics (e.g. accretion efficiency) 
and the mass-transfer process, etc  (see  Tauris \& van den Heuvel 2006; Tauris 2011).
In addition, it has been believed that the accretion-induced collapse (AIC) of  oxygen-neon white dwarfs (ONe WDs)  
will collapse into NSs via electron-capture reactions by Mg and Ne, 
providing an alternative way to form MSPs  
(e.g. Ivanova et al. 2008; Hurley et al. 2010;  Chen et al. 2011, 2023; 
Tauris et al. 2013; Freire \& Tauris 2014;  Guo et al. 2021).

With the development of radio astronomy, more and more MSPs are discovered by the 
current pulsar surveys based on large radio telescopes, such as 
the Parkes radio telescope, the Arecibo telescope and the Five-hundred-meter Aperture 
Spherical radio Telescope (FAST), etc (for a recent review, see Han et al. 2021).
Recent observations indicate that 
some more MSPs with wide seperations ($P_{\rm orb}>30\,\rm d$) are being discovered, but their formation way is still highly unclear
(e.g. Cromartie et al. 2016; Bhattacharyya et al. 2019, 2021; Parent et al. 2019; Bondonneau et al. 2020; 
Deneva et al. 2021; Miao et al. 2023; Gautam et al. 2024).  
Tauris \& Savonije (1999) investigated the NS+red-giant (RG) systems to produce MSPs with wide orbits, 
in which the NSs originating from CCSNe accrete matter from RG donors and spun up to millisecond periods,
named as the CCSN channel. However, Tauris \& Savonije (1999)  did not 
provide the parameter space for producing MSPs through this channel, and thereby the rate. 
It is worth noting that Wang, Liu \& Chen (2022, Paper I) recently studied 
the binary computations of ONe WD+RG systems that undergo AIC and then be 
recycled to form MSPs, called the AIC channel (see also  Ablimit 2023). 
In Paper I, we suggested that the AIC channel could be a viable way to produce MSPs with wide orbits.

In this article, 
we aim to study the formation of MSPs with long orbital periods through 
the CCSN channel  in a systematic way, and then
compare the results with those from the AIC channel.
In Section 2, we introduce the numerical code  and methods for binary evolution computations of NS+RG systems,
and give the corresponding results in Section 3.
In Section 4, we present  the binary population synthesis (BPS) methods, and provide the rate of MSPs with wide orbits.
Finally,  a discussion is shown in Section 5 and a summary in Section 6.

\section{Numerical  code and methods}

In the CCSN channel, the mass donor is a RG star that transfers H-rich matter
and angular momentum  onto the surface of the NS, finally forming MSPs with wide orbits.
We compute the evolution of NS+RG systems until the formation of binary MSPs
by employing the Eggleton stellar evolution code  (see Eggleton 1971, 1972, 1973; 
Han, Podsiadlowski \& Eggleton 1994; Pols et al.\ 1995, 1998).
The basic input physics and initial setup for this code are similar to those in Paper I,
including the convective overshooting parameter, the mass-transfer process and the NS mass-growth rate, etc.
The ratio between mixing length and local pressure scale height is set to be 2.0, and the convective overshooting parameter  to be 0.12, 
approximately corresponding to an overshooting length of $\sim$0.25 pressure scale heights (see Pols et al.\ 1997). 
A classic Population I composition for the initial main-sequence models is adopted with H
abundance $X=0.70$, He abundance $Y=0.28$, and metallicity $Z=0.02$. 
In the stellar models, the number of meshpoints is set to be 399. If we set the number of meshpoints to be 199 and 599 for the 
evolution of NS+RG systems,
we found that there is no significant difference in the results. In addition, we did some tests for the influence of the temporal resolution 
on our results. We also found that there is almost no  difference in our 
results if we change the temporal resolution (for details see Appendix A). 

During the Roche-lobe overflow (RLOF), we use an integrated  prescription for the mass-transfer process with RG donors, 
i.e. an adiabatic power-law mass-transfer  assumption  (see Ge et al. 2010),  as follows:
\begin{equation}
\dot{M}_{\rm 2}=-\frac{2\pi R_{\rm L}^{\rm 3}}{GM_{\rm 2}}f(q)
\int_{\rm \phi_{\rm L}}^{\rm \phi_{\rm s}}\Gamma^{\rm 1/2}(\frac{2}{\Gamma+1})^{\rm \frac{\Gamma+1}{2(\Gamma-1)}}(\rho P)^{\rm 1/2} \rm d\phi,
\end{equation}
in which $\dot{M}_{\rm 2}$ is the mass-transfer rate,  $M_{\rm 2}$ is the mass of the donor, 
$G$ is the gravitational constant, $P$ is the local gas pressure,  $R_{\rm L}$ is the radius of the Roche-lobe, 
$\rho$ is the local gas density, $\phi_{\rm L}$ is the Roche-lobe potential energy 
$\phi_{\rm s}$ is the stellar surface potential energy, and $\Gamma$ is the adiabatic index.
We made more discussions for this mass-transfer process in Section 5.

In our calculations, we do not compute the structure of the NS and suppose it as a point mass, 
in which we set the initial mass of the NS ($M_{\rm NS}^{\rm i}$) as  $1.4\,\rm \rm M_{\odot}$ 
(a canonical NS mass; see van den Heuvel 2009; Chen \& Liu  2013). 
We follow the growth and spin-up of the NS, along with the mass loss of the 
RG donor and its subsequent evolution to the WD stage. 
We adopt the prescription of Tauris et al. (2013) to obtain the mass-growth rate of the NS:
\begin{equation}
\dot{M}_{\rm NS}=(|\dot{M}_{\rm 2}|-\max[|\dot{M}_{\rm 2}|-\dot{M}_{\rm Edd},0])\cdot k_{\rm def}\cdot e_{\rm acc},
\end{equation}
in which $\dot{M}_{\rm Edd}$ is the Eddington accretion rate of the NS,
$k_{\rm def}$ is the mass ratio of the gravitational mass to the remaining mass of the accreted matter, and
$e_{\rm acc}$ is the mass fraction of the transferred matter  to the rest on the NS.
For H accretion onto the NS, $\dot{M}_{\rm Edd}$ is given by Tauris et al. (2013):
\begin{equation}
\dot{M}_{\rm Edd}=4.6\times 10^{\rm -8} {\rm M_{\odot}\,\rm yr^{\rm -1}} \cdot M^{\rm -1/3}_{\rm NS}\cdot \frac{1}{1+X_{\rm H}},
\end{equation}
where $X_{\rm H}$ is the H mass fraction of the accreted matter.
In this work,  $k_{\rm def}$ and $e_{\rm acc}$ are combined into a free parameter $k_{\rm def}\cdot e_{\rm acc}$ 
that represents the mass retention efficiency of the NS. 
We set $k_{\rm def}\cdot e_{\rm acc}$ to be 0.35 on the basis of the increased evidence of the inefficient 
accretion for LMXBs even though $|\dot{M}_{\rm 2}|< \dot{M}_{\rm Edd}$ 
(see, e.g. Jacoby et al. 2005; Antoniadis et al. 2012; Tauris et al. 2013; Ablimit \& Li 2015).  
\footnote{The inefficient accretion may arise from some possible mechanisms, 
e.g. direct atmosphere irradiation of the donor star from the pulsar,
 the instabilities of accretion disc, and the propeller effects, etc
 (see Tauris et al. 2013; see also van Paradijs 1996; Dubus et al. 1999).}
Accordingly,  the NS mass-growth rate can be expressed as:
 \begin{equation}
\dot{M}_{\rm NS}=\left\{
 \begin{array}{ll}
 0.35\dot{M}_{\rm 2}, & |\dot{M}_{\rm 2}|< \dot{M}_{\rm Edd},\\
0.35\dot{M}_{\rm Edd}, & |\dot{M}_{\rm 2}|\geq \dot{M}_{\rm Edd}.
\end{array}\right.
\end{equation}
During the NS mass-growth process,  
we suppose that the
excess material ($|\dot{M}_{\rm 2}|-\dot{M}_{\rm NS}$) is ejected from the vicinity of the NS,
taking away the specific orbital angular momentum of the accreting NS.  For the mass-loss process, we adopted the
isotropic re-emission mechanism, in which the excess material was transferred through an accretion disk to the vicinity of the NS, 
from where it is subsequently ejected in the form of  a rapid isotropic wind
(see, e.g. Soberman, Phinney \& van den Heuvel 1997; Tauris \& Savonije 1999). 
The NS mass-growth rate used in this work is widely adopted in previous studies, see e.g. Ablimit \& Li (2015),
Liu et al. (2018), Tang, Liu \& Wang (2019) and Chen et al. (2021), etc.

The accreted matter will recycle the NS that 
may undergo the spin-up process (e.g. Li et al. 2021). 
We calculate the minimum spin period of the recycled NS before the spin-down process
based on  the prescription of Tauris, Langer \& Kramer (2012):
\begin{equation}
P_{\rm spin}^{\rm min}\sim 0.34\times(\Delta M_{\rm NS}/\rm M_{\odot})^{\rm -3/4},
\end{equation}
in which $\Delta M_{\rm NS}$ is the mass of the accreted matter onto the NS,
and $P_{\rm spin}^{\rm min}$ is in units of ms.
Here, we neglect  the initial  spin velocity of the NS and the gravitational 
binding energy of the accreted matter onto the NS. 
It is worth noting that  Guo et al. (2023) recently explained  the formation of the observed isolated mildly recycled pulsars 
with velocity $\rm < 360\,km\,s^{-1}$ on the basis of equation (5).

\section{Binary evolution results}
In order to form MSPs with wide orbits, we performed a large number of detailed binary evolution computations of 
NS+RG systems that interact through mass-transfer, and thus we obtain a dense grid of binaries. 
In Table 1, we show the main evolutionary features of some selected  NS+RG systems that can form recycled  MSPs.
In this table,  we explored the effect of different initial orbital periods (see sets $1-10$) and 
initial donor masses (see sets $11-15$) on the final results.

\begin{table*}
\begin{center}
\caption{Summary of some typical  NS+RG systems that can evolve into MSPs
with different initial donor masses and initial orbital periods, in which we set $M_{\rm NS}^{\rm i}=1.4\,\rm \rm M_{\odot}$.
The columns (from left to right):  the initial donor mass, the initial orbital period; the stellar age at the beginning of RLOF;
the time-scale that the binary appears as a LMXB; 
the final NS mass, the final donor mass (i.e. the WD mass), and the final orbital period; 
and the minimum spin period of the NS on the basis of equation (5).}
\begin{tabular}{ccccccccccccccc}
\hline\hline
Set & $M_2^{\rm i}$ & $\log P_{\rm orb}^{\rm i}$ &  $t_{\rm RLOF}$ & 
$\bigtriangleup t_{\rm LMXB}$ &  $M_{\rm NS}^{\rm f}$ & $M_2^{\rm f}$ & $\log P_{\rm orb}^{\rm f}$  &  $P_{\rm spin}^{\rm min}$\\
 &   ($\rm M_{\odot}$)  & (d)  &  (Gyr)  & (Myr)  & ($\rm M_{\odot}$) & ($\rm M_{\odot}$)  & (d) & (ms)\\
\hline
 1& 1.4 & $0.4$ &  3.438  & 233 &1.7796 &0.3000 &1.7569   & 0.70  \\
  2& 1.4 & $0.6$ & 3.494  & 127 &1.7600 &0.3134 &1.9055 &    0.73  \\
  3& 1.4 & $0.8$ & 3.536  &  102 &1.7289 &0.3276 &2.0531 &    0.78  \\
  4& 1.4 &$1.0$ &  3.568 &  65  &1.7144 &0.3434 &2.2011 &    0.81 \\
  5& 1.4 &$1.2$ &  3.600 &  34 & 1.6395 &0.3607 &2.3475&    0.99  \\
  6& 1.4 &$1.4$ &   3.618&  17 &1.5733 &0.3808 &2.4980 &   1.27  \\
  7& 1.4 &$1.6$ &   3.621  & 15 &1.5189 &0.4036 &2.6452 &    1.68 \\
  8& 1.4 &$1.8$ &    3.627 & 10 & 1.4825 &0.4295 &2.7850 &    2.21 \\
  9& 1.4 &$2.0$ &   3.631 &  7 & 1.4579 &0.4594 &2.9174 &    2.88 \\
  10&1.4 & $2.2$ &  3.634 & 5 &1.4407 &0.4955 &3.0402 &    3.75  \\
  &&&&&&&&\\
  11&1.1& $1.4$ &   8.416 & 26  &1.5956& 0.3659 &2.3986 &    1.16\\
  12&1.3& $1.4$ &  4.684  &23&1.5828 &0.3765 &2.4723 &    1.22\\
  13&1.5& $1.4$ &  2.874  &21&1.5618 &0.3848 &2.5165 &    1.33\\
  14&1.7& $1.4$ &  1.939  &17&1.5361 &0.3921 &2.5348 &    1.52\\
  15&1.9& $1.4$ &   1.391 &13&1.5075 &0.3994 &2.5316 &    1.81\\
\hline
\end{tabular}
\end{center}
\end{table*}

\subsection{A representative example for binary evolution}

\begin{figure*}
\begin{center}
\epsfig{file=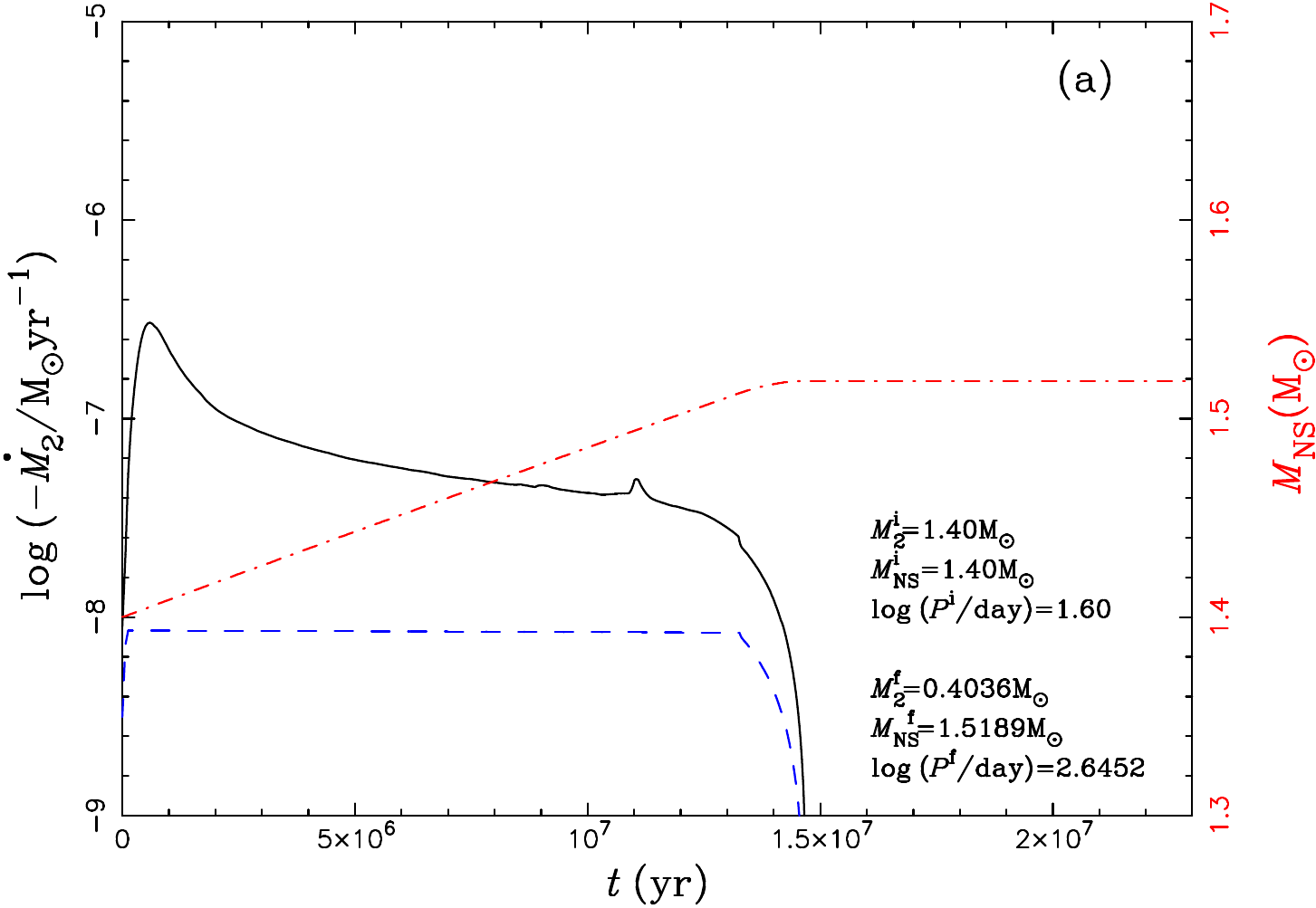,angle=270,width=9.5cm}\ \
\epsfig{file=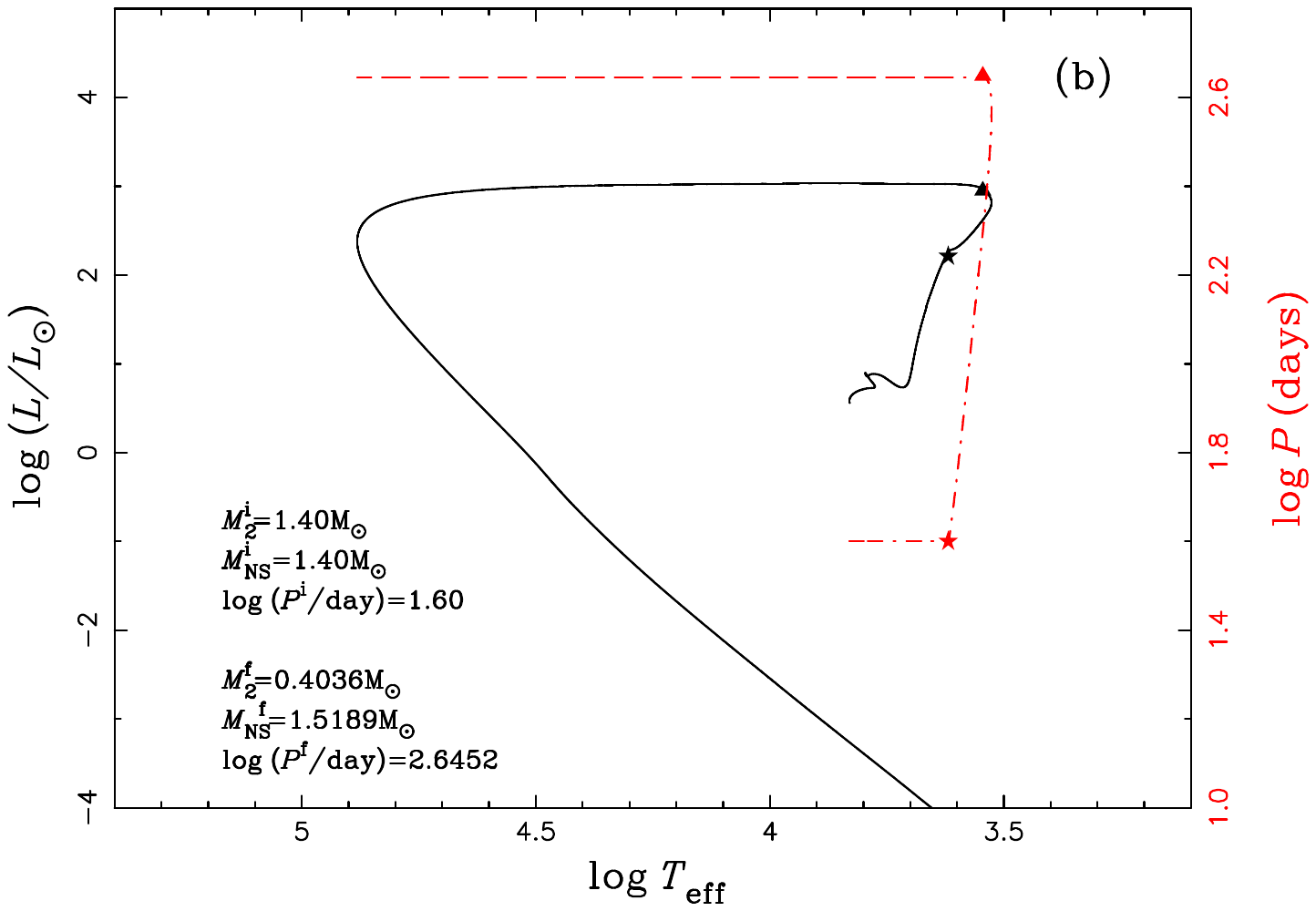,angle=270,width=10.5cm}
  \caption{A representative example for the evolution of a NS$+$RG system until the formation of a recycled MSP,  in which
  ($M_2^{\rm i}$, $M_{\rm NS}^{\rm i}$, $\log (P_{\rm orb}^{\rm i}/{\rm d})$)
$=$ (1.4, 1.4, $1.6$) (see set 7 in Table 1).
Panel (a): the evolution of the $M_{\rm NS}$  (red  dash-dotted line), $\dot{M}_{\rm 2}$ (black solid line) and $\dot{M}_{\rm NS}$ (blue dashed line) 
as a function of time for the binary evolution calculations. 
Panel (b): the binary orbital period (red dash-dotted line) and the luminosity of the mass donor (black solid line)   as a function of effective temperature.
Asterisks indicate the position where the mass-transfer occurs, whereas triangles  mark the position where the mass-transfer stops.}
 \end{center}
\end{figure*}

Fig.\,1 presents a representative example of the evolution of a NS+RG system that 
forms a wide-orbit MSP finally (see set 7 in Table 1).
The initial binary parameters for this system  are
($M_{\rm NS}^{\rm i}$, $M_2^{\rm i}$, $\log (P_{\rm orb}^{\rm i}/{\rm d})$)
$=$ (1.4, 1.4, 1.6), where $M_{\rm NS}^{\rm i}$, $M_2^{\rm i}$ and $P_{\rm orb}^{\rm i}$ are the initial
mass of the NS, the initial mass of the donor, and the initial orbital period, respectively. 
After about $3.621\,\rm Gyr$, the donor  fills its Roche-lobe owing to the rapid expansion  of itself  
when it evolves to the RG phase, resulting in a case B mass-transfer process defined by Kippenhahn \& Weigert (1967). 
During this stage, the RG donor contains a convective envelope  and  transfers H-rich matter onto the NS, 
leading to a spin-up process for the NS.

The $\dot{M}_{\rm 2}$ becomes higher than $\dot{M}_{\rm Edd}$  soon after the RLOF.
In this case, the NS grows in mass at a rate of $0.35\dot{M}_{\rm Edd}$, and
the majority of the transferred matter is blown away from the system 
at a rate of ($|\dot{M}_{\rm 2}|-0.35\dot{M}_{\rm Edd}$).
During the mass-transfer process, the binary shows as a LMXB, lasting for about $15\,\rm Myr$.
At the end of the mass-transfer process, 
the spin period  of the pulsar
will  approach about $1.68\,\rm ms$ before the spin-down process.
After that, the RG donor gradually evolves to a He WD after the exhaustion of its H-shell. 
The binary forms a long orbital MSP finally, consisting of  
 a $0.4036\,\rm M_{\odot}$ He WD and a $1.5189\,\rm M_{\odot}$ pulsar  with an orbital period of  about $442\,\rm d$.

\subsection{Initial parameters of NS+RG systems}

\begin{figure}
\begin{center}
\epsfig{file=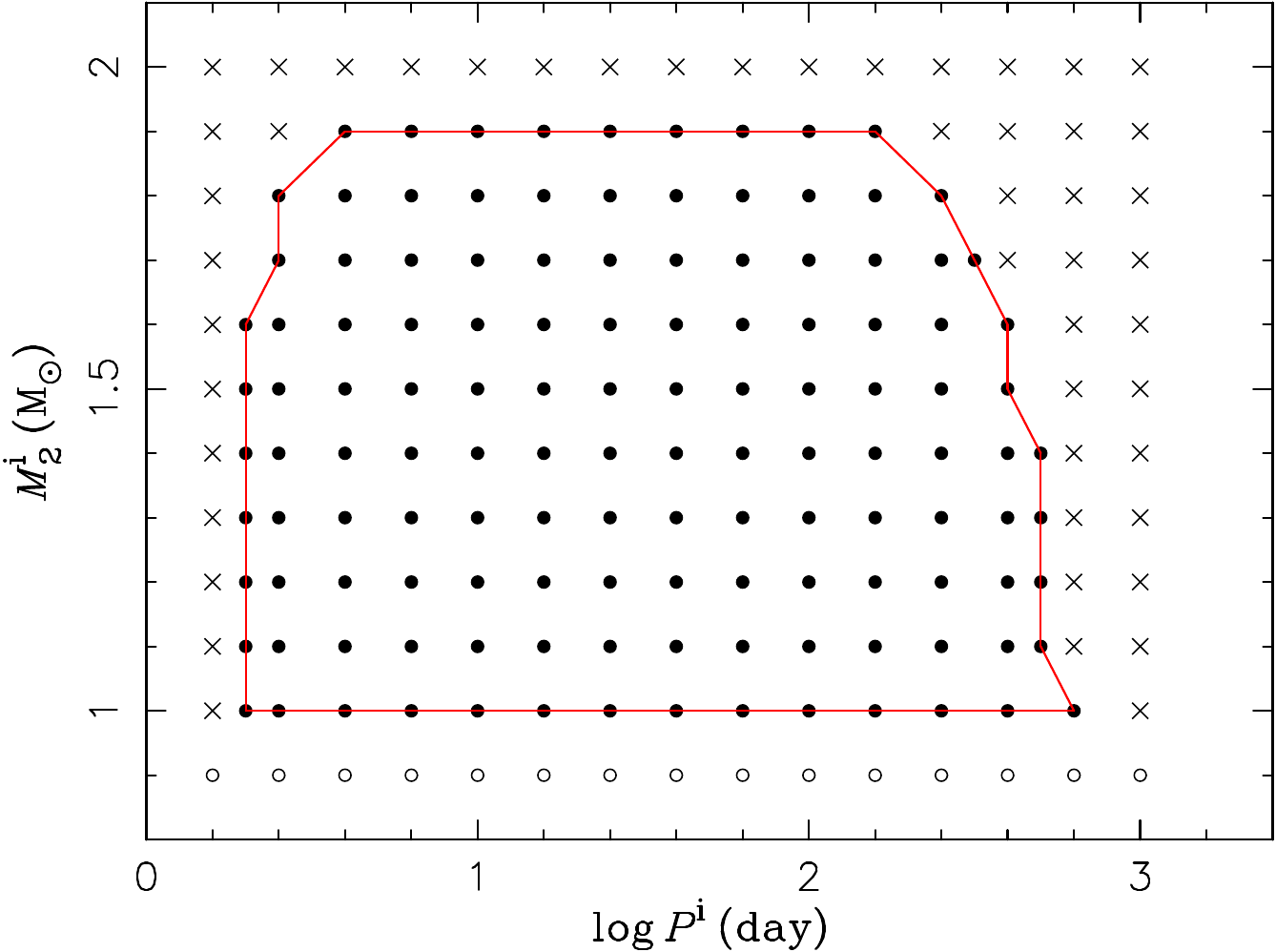,angle=270,width=10.5cm} 
\caption{Initial parameter region of NS$+$RG systems that eventually produce wide orbit MSPs 
in the $\log P^{\rm i}-M^{\rm i}_2$ plane, in which we set  ${M}^{\rm i}_{\rm NS}=1.4\,\rm M_{\odot}$.
The filled circles indicate systems that lead to the formation of MSPs. 
The crosses denote systems that will not form MSPs through the CCSN channel,
and the open circles are those that the mass donors have stellar age larger than 
the Hubble time before filling their Roche-lobes. }
\end{center}
\end{figure}

Fig.\,2 shows the initial parameter space of NS$+$RG systems that eventually form MSPs in 
the $\log P^{\rm i}-M^{\rm i}_2$ plane with ${M}^{\rm i}_{\rm NS}=1.4\,\rm M_{\odot}$, 
in which $P^{\rm i}$ is the initial orbital period of the  NS+RG system and $M^{\rm i}_2$ is the initial mass of the RG donor. 
In order to form MSPs with wide orbits, the NS$+$RG systems should have  initial orbital periods of $\sim2-600$\,d and 
have RG companions with initial masses of $\sim1.0-1.9\,\rm M_{\odot}$. 
The time-scales that the NS+RG systems appear to be LMXBs 
range from $\sim1-100\,\rm Myr$.  In Fig. 2, most of NS$+$RG systems will evolve to NS$+$He WD systems. However,
some of NS$+$RG systems with \textbf{high-mass} RG donors and long orbital periods  
will create hot subdwarfs prior to the WD stage, in which the hot subdwarfs will form CO WDs finally (e.g. sets 9 and 10 in Table 1).

The upper boundary of the initial parameter space in Fig.\,2 is mainly constrained by a high
$\dot{M}_{\rm 2}$ due to a large mass-ratio between the RG star
and the NS, resulting in the formation of a common-envelope (CE).
Systems below the lower boundary have RG donors with stellar age larger than 
the Hubble time when they fill their Roche-lobes. 
The left boundary is determined by the shortest orbital periods when the mass donors
fill their Roche-lobes at the bottom of the RG phase, whereas the right boundary 
are constrained by the condition that the mass donors fill their Roche-lobes at the top of the RG phase.

\subsection{Resulting MSPs}

\begin{table}
\begin{center}
\caption{The relevant parameters of 44 observed MSPs with $P_{\rm spin}<30\,{\rm ms}$ 
detected in the Galactic disk, which have WD companions with wide orbits ($P_{\rm orb}>30\,\rm d$). 
 The observed data are taken from the ATNF Pulsar Catalogue in January 2024 
 (see http://www.atnf.csiro.au/research/pulsar/psrcat; see Manchester et al. 2005). 
The median WD masses are calculated by assuming an orbital inclination angle of $60^{\circ}$
and a pulsar mass of $1.35\,\rm M_{\odot}$. The error bars of the WD masses for the lower limit 
correspond to an inclination angle of $90^{\circ}$, and  the upper limit marks a 90\% probability. }
\begin{tabular}{ccccccccc}
\hline \hline
 $\rm No.$ & $\rm Pulsars$ & $P_{\rm spin}$ &  $\dot{P}_{\rm spin}$            & $P_{\rm orb}$ & $M_{\rm WD}$\\
                  &                        &               (ms)     &                                                  &  (d)   &   ($\rm M_{\odot}$) \\
\hline
1     & $\rm J0203-0150$       &  5.17          &    1.4e-20              &  50.0      &  $0.14^{+0.16}_{-0.02}$ \\   
2     & $\rm J0214+5222$      & 24.58         &    3.0e-19              &  512.0    & $0.48^{+0.70}_{-0.07}$ \\   
3     & $\rm J0407+1607$      & 25.70         &     7.9e-20             &  669.1    &  $0.22^{+0.27}_{-0.03}$\\     
4     & $\rm J0605+3757$      & 2.73           &      4.8e-21            &  55.7      & $0.32^{+0.24}_{-0.03}$\\       
5     &  $\rm J0614-3329$      & 3.15          &        1.8e-20           &  53.6      &  $0.32^{+0.42}_{-0.05}$\\      
6     & $\rm J0732+2314 $     &  4.09          &       6.0e-21           &  30.2      &     $ 0.17^{+0.20}_{-0.02}$ \\   
7     & $\rm  J0921-5202$     &  9.68           &      1.7e-20            &  38.2      &     $ 0.27^{+0.34}_{-0.04}$ \\   
8     & $\rm J1012-4235 $     &  3.10           &       *                     &  38.0      &     $ 0.31^{+0.40}_{-0.05}$ \\   
9     & $\rm J1125-5825$     & 3.10              &      6.1e-20           & 76.4       &     $0.31^{+0.40}_{-0.05}$\\     
10   & $\rm  J1146-6610$    & 3.72              &      0.8e-20           &  62.8      &     $ 0.23^{+0.28}_{-0.03}$ \\   
11   & $\rm  J1312+0051$   &  4.23             &      1.8e-20           &    38.5     &     $ 0.21^{+0.24}_{-0.03}$ \\   
12   & $\rm  J1421-4409$    &    6.39           &       1.2e-20          &      30.7   &     $ 0.21^{+0.24}_{-0.03}$ \\   
13   & $\rm J1455-3330$    &  7.99              &      2.4e-20           &76.2         &    $0.30^{+0.38}_{-0.04}$\\      
14   & $\rm J1529-3828$    &  8.49              &       2.7e-20          &119.7        &    $0.19^{+0.22}_{-0.03}$\\      
15  & $\rm J1536-4948$    &  3.08               &        2.1e-20         & 62.1         &     $ 0.32^{+0.42}_{-0.05}$\\      
16  & $\rm J1623-2631$    &  11.08             &        6.7e-19         & 191.4        &    $0.33^{+0.42}_{-0.05}$\\     
17  & $\rm J1640+2224$   &  3.16              &          2.8e-21        &175.5        &     $0.29^{+0.37}_{-0.04}$\\     
18  & $\rm J1643-1224$    &  4.62             &           1.8e-20        &147.0        &     $0.14^{+0.15}_{-0.02}$\\      
19  & $\rm J1708-3506$   &  4.51              &          1.1e-20         &149.1         &     $ 0.19^{+0.22}_{-0.03}$\\     
20  & $\rm J1713+0747$  &4.57                &           8.5e-21        &67.8          &     $0.32^{+0.42}_{-0.05}$\\      
21  & $\rm J1751-2857$   &3.91                &           1.1e-20        &110.7       &      $ 0.23^{+0.27}_{-0.03}$\\    
22  & $\rm  J1806+2819$    &15.08           &            3.8e-20       &  43.9       &      $ 0.29^{+0.36}_{-0.04}$ \\   
23  & $\rm  J1824-0621$    & 3.23             &            9.1e-21       &   100.9      &    $ 0.34^{+0.45}_{-0.05}$ \\   
24  & $\rm J1824+1014$  &4.07                &           5.4e-21        &  82.6          &   $0.31^{+0.40}_{-0.05}$\\    
25  & $\rm J1825-0319$   &4.55                 &           6.8e-21       &52.6           &    $0.21^{+0.24}_{-0.03}$\\    
26  & $\rm  J1828+0625$   & 3.63             &             4.7e-21      &   77.9        &     $ 0.32^{+0.41}_{-0.05}$ \\   
27  & $\rm J1844+0115$   &4.19               &             1.1e-20      &50.6          &      $0.16^{+0.18}_{-0.02}$\\      
28  & $\rm J1850+0124$  &3.56                &           1.1e-20        &84.9          &      $0.29^{+0.36}_{-0.04}$\\     
29  & $\rm J1853+1303$  &4.09                &           8.7e-21         &115.7       &      $ 0.28^{+0.35}_{-0.04}$\\    
30  & $\rm J1855-1436$   &3.59                &          1.1e-20          &61.5         &      $ 0.31^{+0.40}_{-0.05}$\\     
31  & $\rm J1858-2216 $    & 2.38             &            3.8e-21        &  46.1       &     $ 0.25^{+0.30}_{-0.04}$ \\   
32  & $\rm J1908+0128$    &  4.70            &              5.3e-20      &    84.2     &     $ 0.34^{+0.44}_{-0.05}$ \\   
33  & $\rm J1910+1256$  &4.98                &           9.7e-21          &58.5        &     $ 0.22^{+0.27}_{-0.03}$\\     
34  & $\rm J1913+0618$  &5.03                 &           9.6e-21         &67.7        &      $ 0.33^{+0.43}_{-0.05}$\\     
35  & $\rm J1921+1929$    &2.65               &             3.8e-20       &   39.6      &     $ 0.30^{+0.37}_{-0.04}$ \\   
36  & $\rm J1930+2441$  &5.77                 &             8.7e-21      &76.4          &     $ 0.27^{+0.33}_{-0.04}$\\     
37  & $\rm J1935+1726$ &4.20                  &           *                  &90.8          &     $ 0.26^{+0.31}_{-0.04}$\\     
38  & $\rm J1955+2908$ &6.13                  &           3.0e-20        & 117.3      &      $  0.21^{+0.25}_{-0.03}$\\    
39  & $\rm J2019+2425$ &3.93                   &           7.0e-21       & 76.5        &      $0.36^{+0.49}_{-0.06}$\\     
40  & $\rm J2033+1734$ &5.95                   &           1.1e-20       &56.3         &      $0.22^{+0.26}_{-0.03}$\\     
41  & $\rm J2042+0246$ &4.53                  &           1.4e-20        &77.2          &      $ 0.22^{+0.26}_{-0.03}$\\     
42  & $\rm J2229+2643$ &2.98                 &           1.5e-21         &93.0           &     $0.14^{+0.16}_{-0.02}$\\      
43  & $\rm  J2234+0611$ & 3.58                &            1.2e-20       &   32.0         &     $ 0.22^{+0.26}_{-0.03}$ \\   
44  & $\rm J2302+4442$ &5.19                  &           1.4e-20        &  125.9        &     $ 0.34^{+0.45}_{-0.05}$\\     
\hline \label{1}
\end{tabular}
\end{center}
\end{table}

\begin{figure}
\begin{center}
\epsfig{file=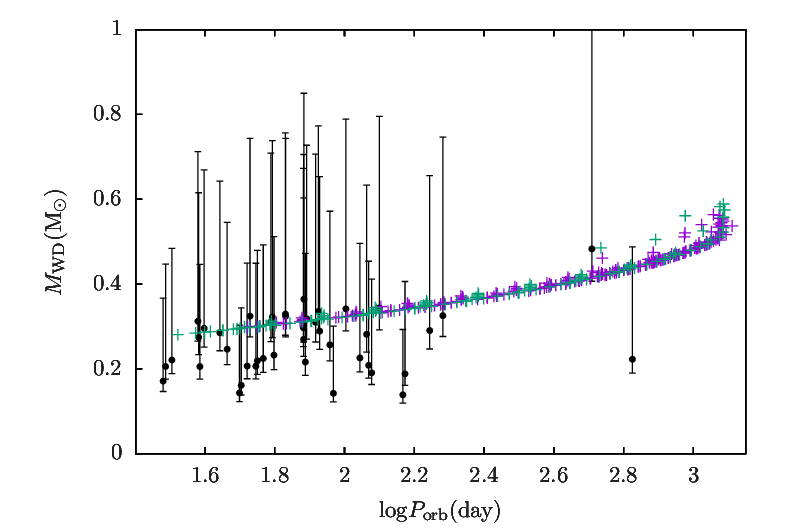,angle=0,width=11.5cm}
 \caption{The WD companion masses of the resulting MSPs formed via  the CCSN channel as a function of orbital periods (see green crosses).
For a comparison, we also show the results for the AIC channel (see purple crosses; see Paper I).
The error bars represent the 44 observed MSPs listed in Table\,2.}
  \end{center}
\end{figure}

In Table 2, we listed some relevant parameters of 44 observed MSPs  with spin periods $<30\,{\rm ms}$ detected in the Galactic disk, 
which have WD companions with orbital periods $>$30\,d.
Fig. 3 shows the resulting binary MSPs formed via the CCSN channel in the final ${M}_{\rm WD}-P_{\rm orb}$ diagram. 
In this figure,  the final binary MSPs follow 
the well-known  relation between  ${M}_{\rm WD}$ and $P_{\rm orb}$,
which can be understood by a relation between the giant's radius and the mass of  its  degenerate core 
(see, e.g. Refsdal \& Weigert 1971;  Rappaport et al. 1995; Tauris \& Savonije 1999; Podsiadlowski, Rappaport \& Pfahl 2002).
From this figure, we can see that the CCSN channel can form  binary MSPs with orbital periods in the range of $30-1200$\,d, 
in which the WD companions have masses  ranging from $0.28\,\rm M_{\odot}$ to $0.55\,\rm M_{\odot}$.
Almost all the observed MSPs with wide orbits can be reproduced 
by the CCSN channel in the ${M}_{\rm WD}-P_{\rm orb}$ diagram.
It is worth noting that metallicity can influence the radius of RGs and thus the ${M}_{\rm WD}-P_{\rm orb}$ relation.
If a lower metallicity is adopted, the ${M}_{\rm WD}-P_{\rm orb}$ relation will be higher than that in Fig. 3, i.e.
for a given donor mass the final orbital period at lower metallicity  is lower than that at higher metallicity 
(see, e.g. Tauris \& Savonije 1999; Zhang et al. 2021).
In addition, the AIC channel can form  binary MSPs with orbital periods  ranging from $50$\,d to $1200$\,d.
Compared with the AIC channel, the CCSN channel can produce binary MSPs 
with shorter orbital periods ($\sim$30\,d) that is mainly decided by the left boundary of the initial parameter space in Fig.\,2.

\begin{figure}
\begin{center}
\epsfig{file=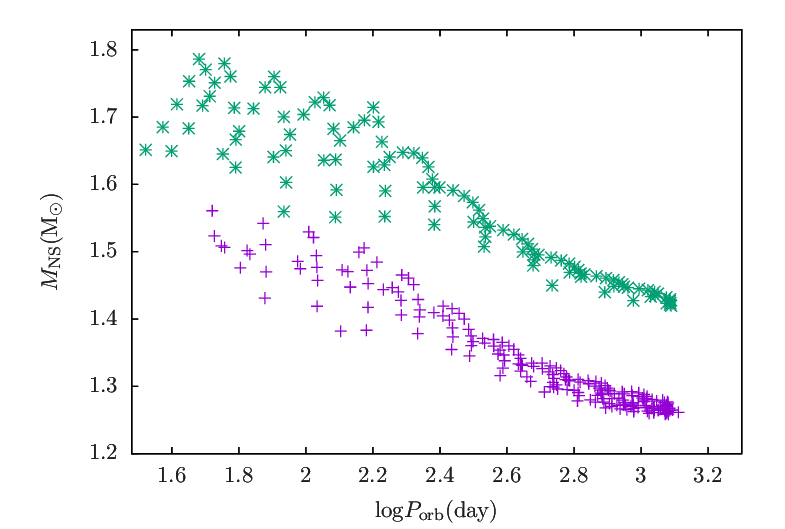,angle=0,width=11.5cm}
 \caption{The NS masses of the resulting MSPs formed through the  CCSN channel as a function of orbital periods (see green crosses).
 For a comparison, we also show the results for the AIC channel (see purple crosses; see Paper I).}
  \end{center}
\end{figure}

Fig. 4 presents the resulting binary MSPs in the ${M}_{\rm NS}-P_{\rm orb}$ diagram.
From this figure, we can see that the final NS masses ranges from $1.42\,\rm M_{\odot}$ to $1.79\,\rm M_{\odot}$, in which
the accreted masses of the recycled pulsars are in the range of $\sim0.02-0.39\,\rm M_{\odot}$. 
For the AIC channel,  the final NS masses are  in the range of  $1.26\,\rm M_{\odot}$ to $1.55\,\rm M_{\odot}$, 
in which the accreted masses  are in the range of $\sim0.01-0.30\,\rm M_{\odot}$ (see Paper I). 
The final NS masses in the AIC channel are lower than those 
from the CCSN channel by assuming the same accretion efficiency of a NS.  
This is because the RG donor in the AIC channel has  already  
lost some of its matter during the pre-AIC evolution, and the NS in the CCSN channel  has a larger initial mass (see van den Heuvel 2009). 
In this figure, we also see that there is an anti-correlation between  ${M}_{\rm NS}$ and $P_{\rm orb}$ for the resulting MSPs,
that is,  MSPs with wide orbits have lower NS masses.
This is because systems with wide orbits have RG donors with larger degenerate cores, 
leading to less layer matter being transferred onto the NSs.
Meanwhile, systems with wide orbits will experience higher mass-transfer process, 
losing too much matter.

\begin{figure}
\begin{center}
\epsfig{file=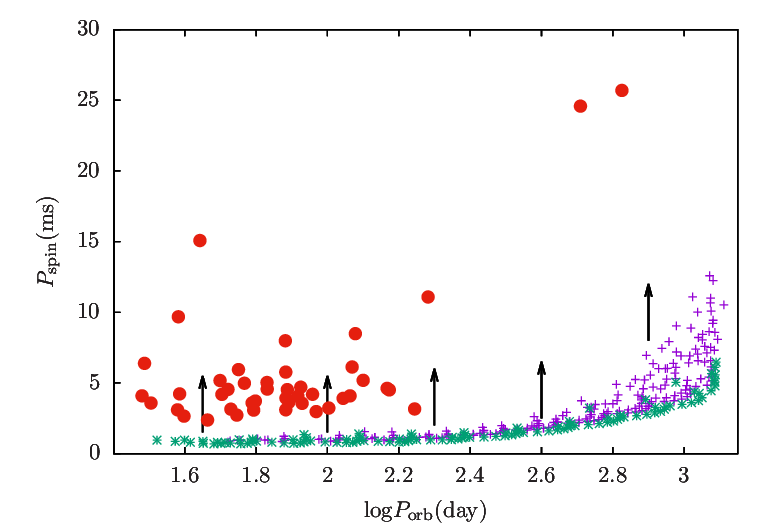,angle=0,width=11.5cm}
 \caption{The Corbet diagram for the resulting MSPs on the basis of the CCSN channel (see green crosses).
For a comparison, we also show the results for the AIC channel (see purple crosses; Paper I).
The filled circles represent the 44 observed MSPs with $P_{\rm spin}<30\,{\rm ms}$  listed in Table\,2.
Note that the theoretical models show a lower limit for the pulsar spin, and
the direction of the arrows indicates that the spin periods will become longer with the spin-down of the pulsars.}
 \end{center}
\end{figure}

Fig. 5 shows the Corbet diagram for the resulting MSPs on the basis of the CCSN channel. 
In the present work, the accreted mass for the recycled pulsars is up to $0.4\,\rm M_{\odot}$, 
thus these MSPs will be fully recycled based on equation (5); pulsars with the accreted mass larger than 
$0.1\,\rm M_{\odot}$ are usually known as fully recycled ones (see Tauris et al. 2013).
The predicted minimum spin periods for these MSPs range from $\sim$1\,ms to 15\,ms before the spin-down process. 
With the spin-down of the pulsars, the spin periods will become longer, indicating 
that the CCSN channel has the potential ability to form the observed MSPs with longer spin periods.
If we set the initial spin period of a pulsar as $2\,{\rm ms}$ (the predicted average spin period 
in minimum) and its $\dot{P}_{\rm spin}=6.26\times10^{-20}$, the  pulsar
needs $1.5-3.0\,{\rm Gyr}$ spin down to $5\,{\rm ms}$ (the average spin period in observations) 
by adopting the value of the magnetic braking index as 3 
(see Fig. 11 of Tauris, Langer \& Kramer 2012).
Compared with the AIC channel (see Paper I), the CCSN channel has the lower predicted minimum spin periods for MSPs. 
This is because the CCSN channel  has the larger accreted masses for the recycled pulsars.
It is worth noting that there is a significant decrease of MSPs with orbital periods $>$200\,d in observations. 
This is mainly due to some observational selection effects for MSPs with wide orbits (for more discussions see Sect. 4.3).

\section{Binary population synthesis}
 
\subsection{BPS methods}

In order to study the Galactic rates of MSPs with wide orbits through the CCSN channel, 
we performed a series of Monte Carlo BPS simulations on the basis of the Hurley's
rapid binary evolution code (see Hurley, Tout \& Pols 2002). 
The basic initial input and assumptions for  Monte Carlo BPS simulations
are  similar to those in Wang et al. (2021),  
including the initial mass-ratio distribution,  the initial orbital separation distribution and  the
initial mass function (IMF), etc (for recent reviews on BPS simulations, see e.g. Han et al. 2020; Chen, Liu \& Han 2024).
The following basic assumptions are adopted:
\begin{enumerate}
\item [(1)]  We assume that all stars are produced in the form of binaries from zero-age main-sequence  (i.e. primordial binaries), in which
a circular orbit  is set. 
\item [(2)]  A  constant mass-ratio distribution is supposed for primordial binaries,  
\begin{equation}
n(q)=1, \hspace{2.cm} 0<q\leq1,
\end{equation}
in which $q=M_{\rm 2, i}/M_{\rm 1, i}$,  where 
$M_{\rm 1, i}$ and $M_{\rm 2, i}$ are the initial masses of the primordial primary and secondary, respectively (see Mazeh et al. 1992; Goldberg \&
Mazeh 1994; Shatsky \& Tokovinin 2002).
\item [(3)]  We supposed the orbital separation distribution as constant in $\log a$ for  primordial binaries with wide orbits,
where $a$ is the orbital separation of the primordial binary. 
\item [(4)]  Since the MSP progenitor 
comes from massive stars, we adopt the IMF of Kroupa (2001) for the primordial primaries, 
in which $M_{\rm 1, i}$ ranges from 0.1\,$M_{\rm \odot}$ to 100\,$M_{\rm \odot}$.
\item [(5)] We adopt a constant star formation rate over the last 15\,Gyr, 
where we assume  that a primordial binary 
with its primary $>0.8\,\rm M_{\odot}$ is produced every year  (see
Han, Podsiadlowski \&  Eggleton 1995; Hurley, Tout \& Pols 2002).  
According to this calibration, we can  obtain  a constant  star formation rate of $\sim$$5\,\rm M_{\odot}{yr}^{-1}$ 
 (see Willems \& Kolb 2004). 
\end{enumerate}

In each simulation, we follow the evolution of a sample of $1\times10^{\rm 7}$ 
binaries from primordial binaries
until the production of NS+RG systems based on a binary evolutionary way (see Sect. 4.2), in which
we did not consider the contribution of  the AIC channel to NS+RG systems.
A MSP with a wide orbit is assumed to be formed once the initial binary parameters of a
NS+RG system are located in the initial parameter space of Fig. 2. 	
It is worth noting that MSPs probably emerge from the evolution of CE  in giant binaries.
Accordingly, we  adopt the standard energy prescription to 
obtain the output of the CE evolution (see Webbink 1984).
As in previous studies (see e.g. Wang et al. 2009, 2021), 
we combine the stellar structure parameter ($\lambda$)  and the CE ejection efficiency ($\alpha_{\rm CE}$) 
into  a free parameter (i.e. $\alpha_{\rm CE}\lambda$). 
In order to explore the effect of different  values of $\alpha_{\rm CE}\lambda$  on the final results,
we set $\alpha_{\rm CE}\lambda=$ 1.0 and 1.5 for a comparison.

\subsection{Evolutionary way to NS+RG systems}

\begin{figure}
\begin{center}
\epsfig{file=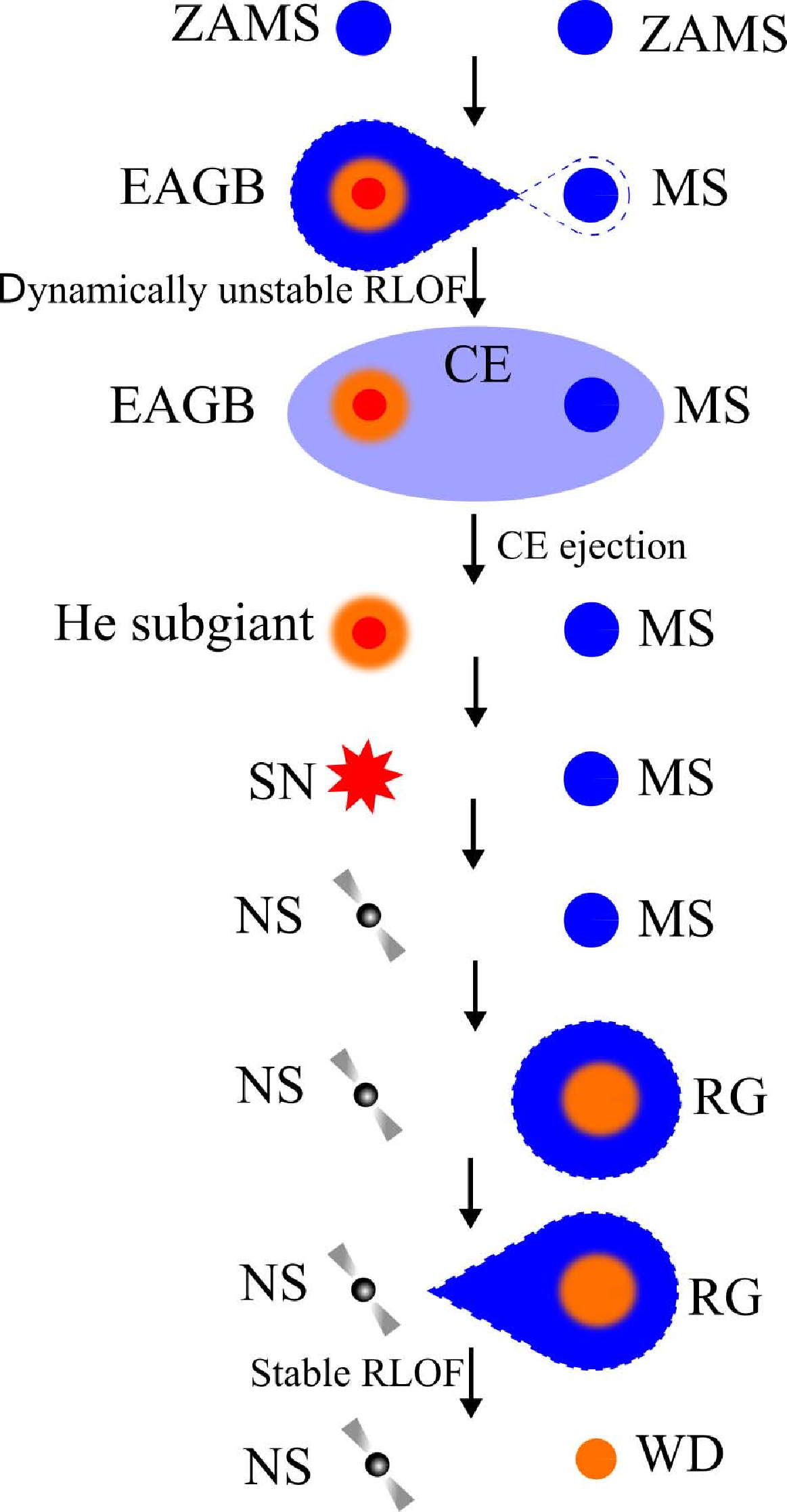,width=7.5cm} \caption{Binary evolutionary way to NS+RG systems that can form MSPs with wide orbits.}
\end{center}
\end{figure}

According to the evolutionary stage of the primordial primary (i.e. the progenitor of the NS) at the beginning of the first RLOF, 
there is one binary evolutionary way that can produce NS+RG systems and then form MSPs with wide orbits (see Fig. 6).
The primordial primary first fills its Roche-lobe at the early asymptotic giant branch (EAGB) stage. 
In this case, the mass-transfer is dynamically unstable due to the large mass-ratio and a CE will be formed.
If the CE can be ejected, the primordial primary becomes a He subgiant star (a He star at the Hertzsprung gap stage; 
see Hurley, Tout \& Pols 2002), and then the binary continues to evolve.
The He subgiant star continues to evolve, leading to the formation of a NS via a CCSN.
A NS+RG system will be formed when the primordial secondary evolves to be a RG star.
When the RG star fills its Roche-lobe, the binary behaves as a LMXB.
After the mass-transfer process, a MSP with a He WD companion will be produced.
In some cases, the mass donors will create hot subdwarfs prior to the WD stage, finally forming CO WDs (see Sect. 3.2).
For this channel, the initial parameters of the primordial binaries for producing wide-orbit MSPs are in the range of 
$M_{\rm 1, i} \sim 10-20\,\rm M_{\rm \odot}$, $q = M_{\rm 2, i}/M_{\rm 1, i} <0.2$, and $P^{\rm i} \sim 600-3800$ days,
where  $q$ and $P^{\rm i}$
are the primordial mass ratio and the initial orbital period, respectively. 

\subsection{BPS results}

\begin{table}
 \begin{center}
 \caption{The estimated Galactic rates and numbers 
 of the resulting MSPs for the CCSN and AIC channels with different values of CE ejection parameters,  in which
 we adopt metallicity $Z=0.02$ and a constant
star-formation rate of $5\,{M}_{\odot}\rm yr^{-1}$ in our Galaxy. 
 Notes: 
 $\alpha_{\rm CE}\lambda$ = CE ejection parameter; 
$\nu_{\rm MSP}$ = Galactic rates of MSPs;
$\rm Number$ = Expected  number of  resulting MSPs in the Galaxy.}
   \begin{tabular}{cccccccccccccc}
\hline \hline
$\rm Channels $ & $\alpha_{\rm CE}\lambda$  & $\nu_{\rm MSP}$  & $\rm Number$ \\
&  &($10^{-5}$\,yr$^{-1}$) & $(10^{\rm 5})$ \\
\hline
$\rm The\,CCSN\,channel$      & $1.0$    & $4.0$      & $4.8$   \\
$\rm The\,CCSN\,channel$      & $1.5$    & $7.1$      & $8.5$    \\
\hline
$\rm The\,AIC\,channel$          & $1.0$    & $ 2.1$      & $2.5$   \\
$\rm The\,AIC\,channel$          & $1.5$    & $3.2$      & $3.8$  \\
\hline
\end{tabular}
\end{center}
\end{table}

Table\,3 presents the estimated Galactic rates and numbers of the resulting MSPs with wide orbits, 
in which we compare the results between 
the  CCSN and AIC channels with different CE ejection parameters. 
For the AIC channel, we obtain the BPS results based on Fig. 3 in Paper I.

\begin{figure}
\begin{center}
\epsfig{file=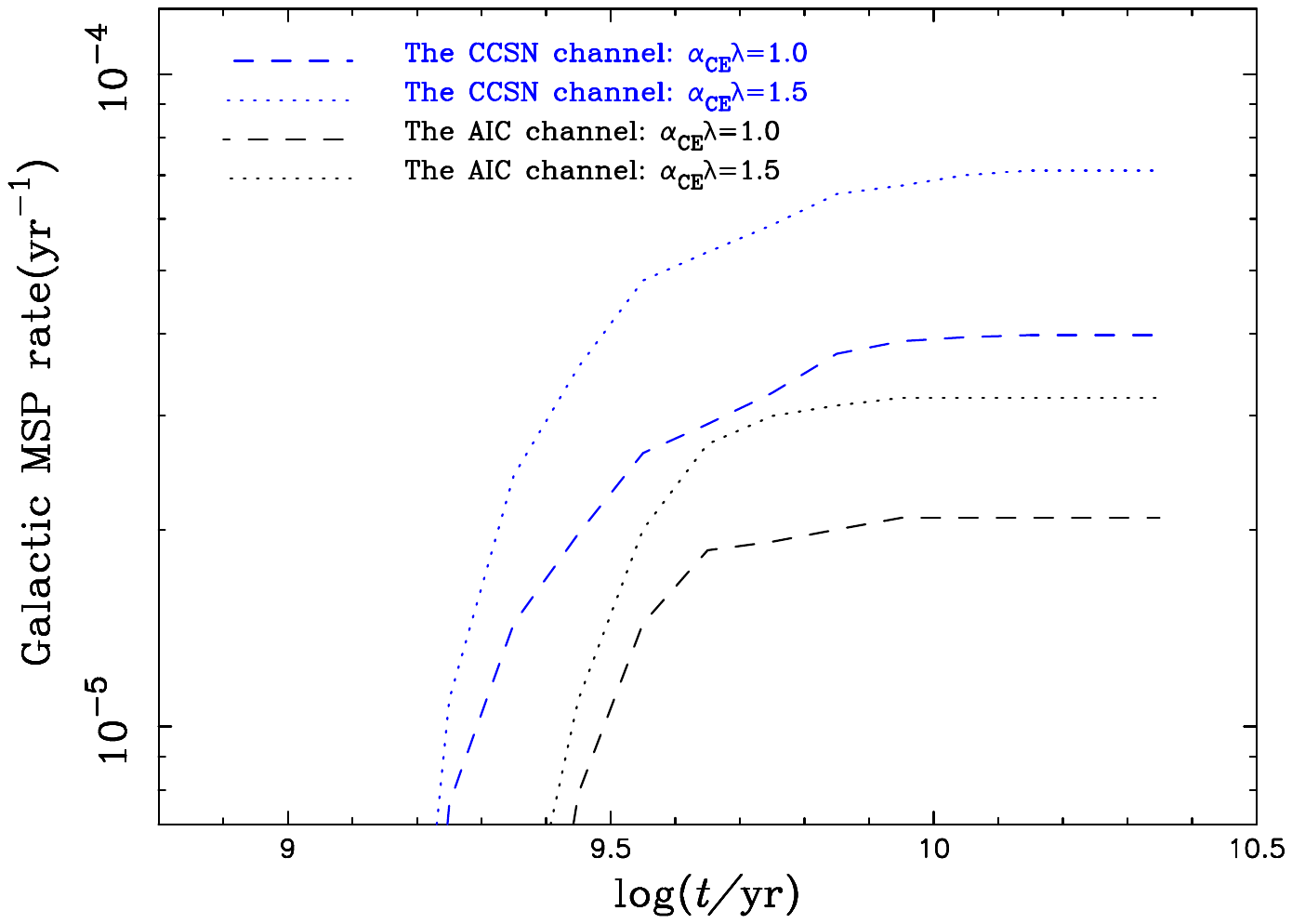,angle=0,width=12cm} 
\caption{Evolution of the Galactic rates  of MSPs from the CCSN channel as a function of time 
by adopting  metallicity $Z=0.02$ and a constant star formation rate of $5~\rm M_{\odot} yr^{\rm -1}$.
For a comparison, we also show the results for the AIC channel.}
\end{center}
\end{figure}

Fig. 7 shows the evolution of the Galactic rates  of MSPs from the CCSN channel
by adopting a constant star formation rate of $5~\rm M_{\odot} yr^{\rm -1}$.
The simulations give the Galactic rates  of MSPs from the CCSN channel to be  $\sim 4.0-7.1\times10^{-5}\,\rm yr^{-1}$,
and the numbers of the resulting MSPs are $\sim 4.8-8.5\times10^{5}$ in the Galaxy.
In Fig. 7, we set  an initial  Population I composition ($Z=0.02$) for the binary evolution computations.
If we adopt a lower value of metallicity, the initial region forming MSPs in Fig.\,2 will move to shorter orbital periods
and lower masses of companions, resulting a lower Galactic rates  of MSPs  (see, e.g. Meng, Chen  \& Han 2009; Wang \& Han 2010). 
It is worth noting that the predicted number of Galactic wide-orbit MSPs is much higher than these in observations.
This is mainly due to some observational selection effects, as follows:
(1) It is hard to observe MSPs with long orbital periods  (see Wang et al. 2020).
(2) We can only detect  these pulsars, whose radiated beams of light orientate themselves towards us.
(3) These observations that only detected 44 wide-orbit MSPs in the Galactic disk (see Table 2) have their only selection effects,
such as the detection limit of radio luminosity and the observational area, etc. 
Therefore, it is hard to directly compare our predictions with these in observations.

In Fig. 7, we also show the results of the AIC channel for a comparison. According to the AIC channel, 
the predicted Galactic rates  of MSPs with wide orbits are $\sim 2.1-3.2\times10^{-5}\,\rm yr^{-1}$,
and the numbers of the resulting MSPs are in the  range of $\sim 2.5-3.8\times10^{5}$ in the Galaxy (see Paper I; Wang 2018).
Compared with the AIC channel, 
the CCSN channel plays a main way to produce MSPs with wide seperations.
From this figure, we can see that
the estimated Galactic rates of MSPs are strongly dependent on the value of the CE ejection 
parameter $\alpha_{\rm CE}\lambda$.
However, the CE ejection 
parameter is still highly uncertain (e.g. Nelemans \& Tout 2005; Ivanova et al. 2013).
Note that 
 $\alpha_{\rm CE}$ and/or $\lambda$ might be lower than 1 as several studies suggested (see, e.g. Dewi \& Tauris 2000; Scherbak \& Fuller 2023). 
If we adopt a lower value of $\alpha_{\rm CE}\lambda$ (e.g. 0.5), the Galactic rates  of MSPs from the CCSN and AIC channels will decrease to 
$\sim 1.7\times10^{-5}\,\rm yr^{-1}$ and $\sim 1.9\times10^{-5}\,\rm yr^{-1}$, respectively.

\section{Discussion}

Previous investigations usually adopted a surface boundary criteria to deal with 
the mass-transfer process once the donor star fills its Roche-lobe, as follows:
\begin{equation}
\dot{M}_{\rm 2}= -C\max[0,{(\frac{R_2}{R_{\rm L}}-1)^{\rm 3}}],
\end{equation}
in which $R_2$ is the donor radius, $R_{\rm L}$ is the radius of its Roche-lobe, and 
$C$ is a dimensionless constant usually assumed to be $\rm 1000\,\rm M_{\odot}\,yr^{\rm -1}$
 (e.g. Han, Tout \& Eggleton 2000; Wang \& Han 2010). 
On the basis of this assumption, the exceeding mass of the donor would be transferred onto the surface of  the accretor  
when the RLOF happens. 
It has been suggested that the constant $C$ in equation (7) is too large for semidetached binaries with giant donors, 
likely overestimating the mass-transfer rate  (see Liu et al. 2019, and references therein).

In this work, by assuming that the state equation for RG donors  follows an adiabatic power law, 
and that the mass outflow is laminar and occurs along the equipotential surface,
we adopted an approximate criterion for the process  of the mass-transfer, see equation (1). 
In this prescription, the $\dot{M}_{\rm 2}$ changes with the local matter states, corresponding to a variable $C$ in equation (7).
Meanwhile, the
$\dot{M}_{\rm 2}$ for RG donors will be lower than that in equation (7), 
leading to a lower mass-loss rate, and thus more matter will be accumulated onto the NS.
It has been argued that the birthrate of type Ia SNe from the semidetached symbiotic channel with RG donors
is relatively low (see, e.g. Li \& van den Heuvel 1997; Han \& Podsiadlowski 2004).
By using this mass-transfer assumption, however, Liu et al. (2019) recently enlarged  
the parameter space for producing  type Ia SNe significantly
based on the semidetached symbiotic channel.
It is worth noting that the integrated mass-transfer assumption for RG donors is still an open question. 
Woods \& Ivanova (2011) argued that a RG star  will not expand adiabatically when 
the local thermal timescale of the superadiabatic outer surface
is comparable with the mass-loss timescale.
This implies that this work might underestimate the mass-transfer rate, 
but  at least we provide an upper limit of the initial parameter space forming binary MSPs with wide seperations.

In the present work, we only consider the formation of binary MSPs  in the Galactic disk. 
In the dense globular clusters, however, MSPs with wide orbits can be formed through
isolated pulsars that have captured WDs  (see, e.g. Verbunt \& Freire 2014).
Meanwhile, the AIC channel could produce newborn pulsars with small kicks, and 
thus this channel can be used to  explain obviously
young  pulsars  in some globular clusters  (see, e.g. Boyles et al. 2011). 
It is worth noting that the AIC channel  may help to solve the observed discrepancy 
between the large rate of MSPs and the small rate of LMXBs in the Galaxy
(see, e.g. Kulkarni \& Narayan 1988; Bailyn \& Grindlay 1990; Tauris et al. 2013).
For a recent review on the formation of MSPs through the AIC channel, see Wang \& Liu (2020).

In the observations, NSs with RG donors can be identified as symbiotic 
X-ray binaries that are a rare class of LMXBs.  They can also show as
ultraluminous X-ray sources during high mass-transfer stage (see, e.g.  Shao \& Li 2015;  Misra et al. 2020).
There are several observed symbiotic X-ray binaries that consist of 
pulsars and RG donors, as follows:
 (1) V2116 Oph  (i.e. GX 1+4) has a pulsar with an orbital period of $\sim$1161\,d,
 in which the pulsar with  a spin period of $\sim$2\,min (see, e.g. 
 Hinkle et al. 2006; I{\l}kiewicz, Miko{\l}ajewska \& Monard 2017).
Hinkle et al. (2006) suggested that the maximum mass of the RG donor is  
about $1.22\,\rm M_{\odot}$ by adopting a NS mass of $1.35\,\rm M_{\odot}$.
(2) V934 Her (i.e. 4U 1700+24) has  a pulsar with an orbital period of $\sim$12\,yr, 
in which the observed properties of the pulsar  are driven by mass accretion from the RG via stellar wind (see, e.g. Hinkle et al. 2019).  
(3) SRGA J181414.6-225604 is an unique case of a symbiotic X-ray system, in which 
the steady wind-accretion leads to faint X-ray emission  as the orbit is wide enough (see De et al. 2023).
(4) IGR J17329-2731 has a highly magnetized NS that accretes material via the stellar  
wind from its giant companion (see Bozzo et al. 2018).  
(5) It has been suggested that IGR J16194-2810 is a symbiotic X-ray binary with an orbital period of $\sim$193\,d, 
consisting of a $1.23^{+0.05}_{-0.03}\,\rm M_{\odot}$ NS and a giant mass of
 $0.99^{+0.02}_{-0.03}\,\rm M_{\odot}$  (see Hinkle et al. 2024; Nagarajan et al. 2024).
In future observations, more  symbiotic X-ray binaries are 
expected to test the theoretical models in this work.

\section{Summary}
In this work, by employing an adiabatic power-law assumption for the mass-transfer  process, 
we performed a large number of complete binary evolution calculations for the formation of MSPs with wide orbits
through the CCSN channel in a systematic way.
We found that 
the CCSN channel can form MSPs with orbital periods ranging from $30\,{\rm d}$ to  $1200\,{\rm d}$, in which 
the masses of the WD companions are  in the range of $0.28-0.55\,\rm M_{\odot}$. 
We also found that there exists an anti-correlation between the final NS mass and orbital period,
and that all the final MSPs through the  CCSN channel  follow  
the correlation between the companion mass and  the orbital period. 
In the observations, NSs with RG donors can be identified as LMXBs lasting for $\sim1-100\,\rm Myr$.
By using a detailed BPS approach, 
we estimate the  numbers of the resulting MSPs  in the  range of $\sim 4.8-8.5\times10^{5}$ in the  Galaxy.
Compared with the AIC channel,
the CCSN channel provides a main way to explain the observed MSPs with wide seperations, especially in the Galactic disk.
More observational identifications and theoretical simulations on MSPs with wide orbits are needed
for our understanding of this class of wide compact systems.

\section*{Acknowledgments}
We acknowledge the anonymous referee for valuable comments 
that help to improve the paper.
This study is supported by the National Natural Science Foundation of China (Nos 12225304, 12073071, 12273105, 12288102, 12090040, 12090043), 
the National Key R\&D Program of China (Nos 2021YFA1600404), 
the Western Light Project of CAS (No. XBZG-ZDSYS-202117), the Youth Innovation Promotion Association CAS (No. 2021058),
the science research grants from the China Manned Space Project (Nos CMS-CSST-2021-A12/A10), 
the Frontier Scientific Research Program of Deep Space Exploration Laboratory (No. 2022-QYKYJH-ZYTS-016), 
the Yunnan Fundamental Research Projects (Nos 202401AV070006, 202101AT070027, 202101AW070003, 202201AW070011, 202201BC070003),
and the Yunnan Revitalization Talent Support Program (Yunling Scholar Project and Young Talent Project).

\section*{Data availability}
The data of the numerical calculations can be available  by contacting BW.

\newpage
\appendix
\section{The temporal resolution}\label{appendix}

In the Eggleton stellar evolution code, in order to prevent the timestep from fluctuating too rapidly,
the ratio between the next timestep and the present timestep is restricted to lie in the range ($\rm DT1$, ${\rm DT2}$),
where the $\rm DT1$ is usually set to be in the range of  $0.8-1.0$ and  $\rm DT2=1.0-1.2$.
If both  $\rm DT1$  and $\rm DT2$ are 1, then the timestep is constant.

In this work, we set $\rm DT1=0.8$ and  $\rm DT2=1.2$ for our binary evolution calculations. 
In this appendix, we made some tests for the influence of the temporal resolution  on our results. 
In our tests, the initial binary parameters are similar to those in Fig. 1, i.e. ($M_{\rm NS}^{\rm i}$, $M_2^{\rm i}$, $\log (P_{\rm orb}^{\rm i}/{\rm d})$)
$=$ (1.4, 1.4, 1.6), but we set $\rm DT2=1.0$ and 1.3 (see Fig. A1). 
From Fig. A1, we can see that there is almost no  difference in our results when we change the values of $\rm DT2$.

\begin{figure*}
\begin{center}
\epsfig{file=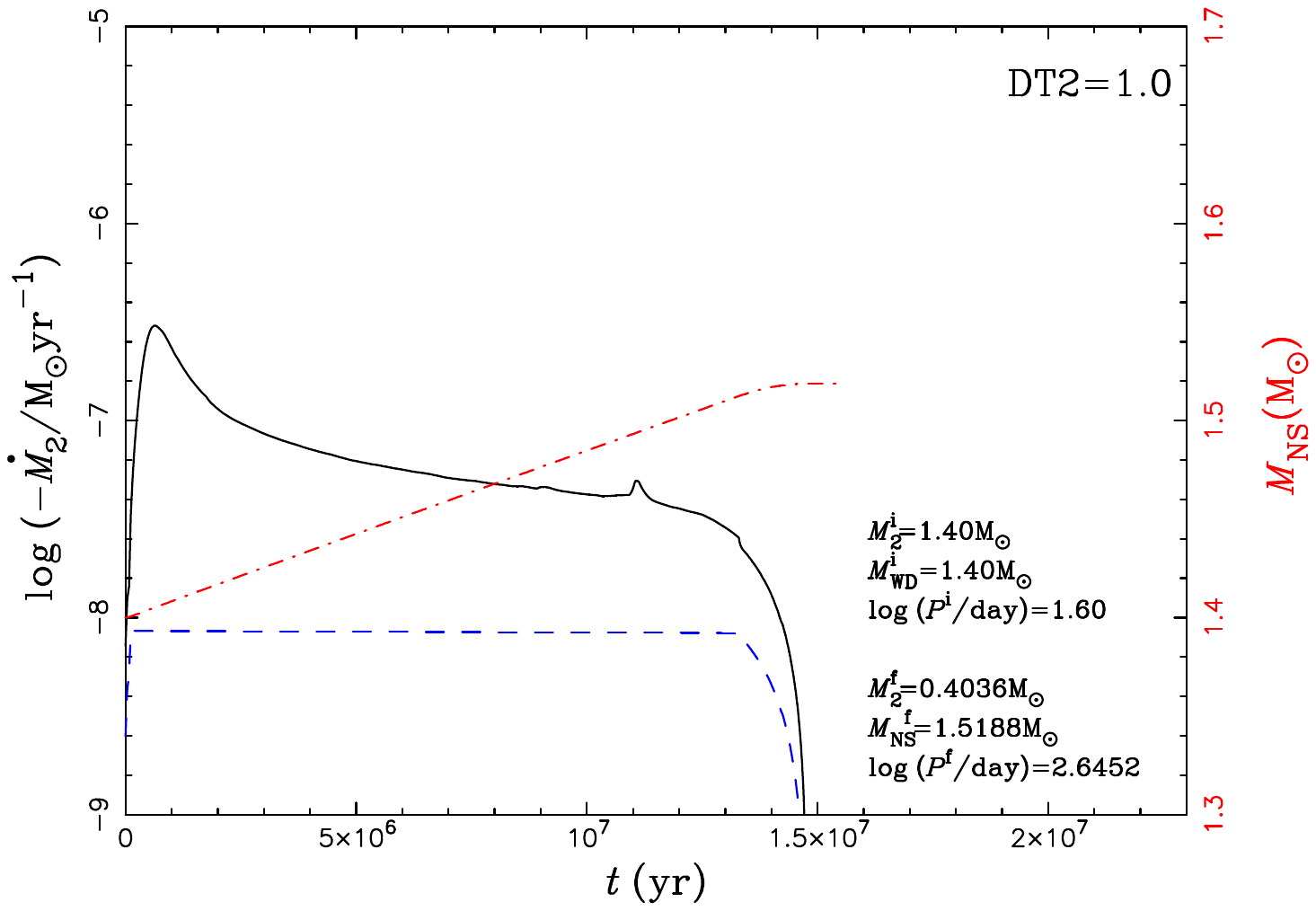,angle=270,width=9.5cm}\ \
\epsfig{file=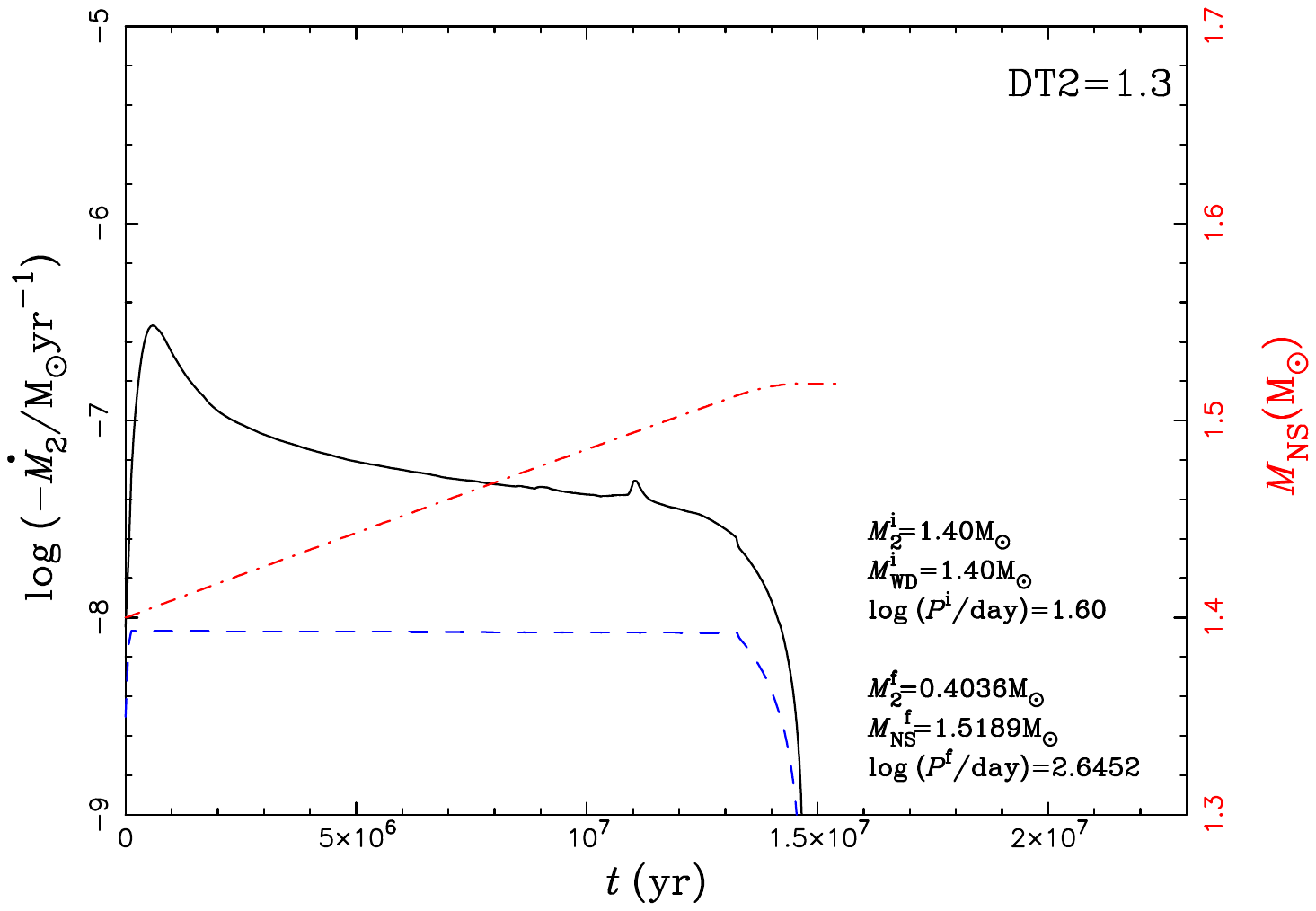,angle=270,width=9.5cm}
  \caption{Similar to Panel (a) of Fig. 1, but for $\rm DT2=1.0$ and 1.3.}
 \end{center}
\end{figure*}

\label{lastpage}
\end{document}